\documentclass[prl,twocolumn,superscriptaddress,floatfix,noshowpacs,10pt,longbibliography]{revtex4-1}%
\usepackage{graphicx,bm,times}
\usepackage{amsmath}
\usepackage{amsfonts}
\usepackage{amssymb}

\begin{document}

\title
{Suppression of electronic correlations by chemical pressure from  FeSe to FeS}

\author{P. Reiss}
\affiliation{Clarendon Laboratory, Department of Physics,
University of Oxford, Parks Road, Oxford OX1 3PU, UK}
\author{M. D. Watson}
\affiliation{Diamond Light Source, Harwell Campus, Didcot, OX11 0DE, UK}
\author{T. K. Kim}
\affiliation{Diamond Light Source, Harwell Campus, Didcot, OX11 0DE, UK}
\author{A. A. Haghighirad}
\affiliation{Clarendon Laboratory, Department of Physics,
University of Oxford, Parks Road, Oxford OX1 3PU, UK}

\author{D. N. Woodruff}
\affiliation{Department of Chemistry, University of Oxford, Inorganic Chemistry Laboratory,
South Parks Road, Oxford, OX1 3QR, United Kingdom}

\author{M. Bruma}
\affiliation{Clarendon Laboratory, Department of Physics,
University of Oxford, Parks Road, Oxford OX1 3PU, UK}

\author{S. J. Clarke}
\affiliation{Department of Chemistry, University of Oxford, Inorganic Chemistry Laboratory,
South Parks Road, Oxford, OX1 3QR, United Kingdom}

\author{A. I. Coldea}
\email[corresponding author:]{amalia.coldea@physics.ox.ac.uk}
\affiliation{Clarendon Laboratory, Department of Physics,
University of Oxford, Parks Road, Oxford OX1 3PU, UK}

\begin{abstract}
Iron-based chalcogenides are complex superconducting systems in which orbitally-dependent electronic correlations  play an important role. Here,  using  high-resolution angle-resolved photoemission spectroscopy, we investigate the effect of these electronic correlations outside the nematic phase in the tetragonal phase of superconducting FeSe$_{1-x}$S$_x$ ($x=0, 0.18, 1$). With increasing sulfur substitution, the Fermi velocities increase significantly and the band renormalizations are suppressed towards a factor of $1.5-2$ for FeS. Furthermore, the chemical pressure leads to an increase in the size of the quasi-two dimensional Fermi surface, compared with that of FeSe, however, it remains smaller than the predicted one from first principle calculations for FeS.
Our results show that the isoelectronic substitution is an effective way to tune electronic correlations in FeSe$_{1-x}$S$_x$, being weakened
for FeS with a lower superconducting transition temperature. This suggests indirectly that electronic correlations could help to promote higher-$T_c$ superconductivity in FeSe.
\end{abstract}
\date{\today}
\maketitle
%%%%%%%%%%%%%%%%%%%%%%%%%%%%%%%%%%%%%%%%%%%%%%%%

Iron-based superconductors offer an interesting playground to explore the competition
of low-energy electronic ground states, such as superconductivity, spin-density wave and nematic states.
These low energy electronic states are strongly influenced by the presence of the different 3$d$ orbitals of Fe,
the Hund's coupling, Coulomb interactions and band filling \cite{Yin2011,deMedici2014}.
The orbitally-selective nature of these interactions often leads to different
bandwidth renormalizations and an unusual relative energy shift of 
various bands with respect to each other and the Fermi level due to the pronounced particle-hole asymmetry of the electronic structure \cite{Fanfarillo2016}.
Iron-chalcogenides are among the most correlated iron-based superconductors,
displaying the largest spread of orbitally-dependent bandwidth renormalization.
The most pronounced renormalization is observed for the band with $d_{xy}$
orbital character, reaching a factor of 17 and being sensitive to
the isoelectronic substitution, as for FeSe$_{1-x}$Te$_x$
\cite{Yi2015_natcomm,Tamai2010}.

FeSe is a unique system in which the role of correlations on nematicity and superconductivity can be explored, in the absence of a competing long-range spin-density wave order. Superconductivity in FeSe around $T_c \approx 9$\,K emerges out of a nematic electronic state, showing strong anisotropy in the electronic and superconducting properties
\cite{Watson2015a,Tanatar2016,Sprau2016_arxiv}.
The origin of this nematic phase below  $T_s \approx 90$~K,
which coincides with  a tetragonal-orthorhombic structural phase transition
\cite{McQueen2009PRL}, is the orbital order that breaks the four-fold rotational symmetry and thus leads to the lifting of
the $d_{xz}/d_{yz}$ orbital degeneracy \cite{Shimojima2014,Watson2015a}. Previous studies in the tetragonal phase
found orbitally-dependent band renormalizations for FeSe
reaching values from 3-4 for the degenerate $d_{xz}/d_{yz}$ bands, to 7-9 for the $d_{xy}$ band \cite{Maletz2014,Watson2015a}.
Furthermore, the strength of electronic correlations manifests by the existence of a lower Hubbard band at large binding
energies, recently detected in FeSe \cite{Watson2017a,Evtushinsky2016_arxiv}.

Superconductivity in bulk FeSe can be strongly enhanced using various tuning parameters,
such as applied physical pressures \cite{Medvedev2009,Terashima2015}, chemical intercalations \cite{BurrardLucas2013}
and {\it in-situ} potassium dosing \cite{Wen2016}.
The enhancement in superconductivity by doping of the surface \cite{Wen2016} was found to be directly linked to the increase of electronic correlations.
However, sulfur substitution in FeSe$_{1-x}$S$_x$ completely suppresses the nematic state for $x \ge 0.18$(1) \cite{Watson2015a,Watson2015c,Coldea2016},
without promoting a high-$T_c$ superconducting phase or stabilizing a magnetic order, in contrast to applied pressure \cite{Medvedev2009,Terashima2015}.
The end member of this series, the tetragonal FeS, with a lower $T_c \approx 4$\,K, is suggested to be less correlated \cite{Man2016,TerashimaFeS},  emphasizing the important role of chemical pressure in tuning electronic ground states and the strength of electronic correlations.

In this paper, we study the suppression of electronic correlations and the changes in band structure as a function of isoelectronic substitution in the tetragonal phase of FeSe$_{1-x}$S$_x$ using high-resolution angle-resolved photoemission spectroscopy (ARPES).
The low temperature Fermi surface of the tetragonal phase with
$x=0.18$ resembles that of FeSe at  high temperatures ($T > T_s$),
and it expands towards FeS, but it does not reach the size predicted by first principle calculations.
The Fermi velocities increase and the band renormalizations decrease significantly with increasing $x$.
At the same time, superconductivity is weakened as the electronic correlations of the $d_{xz}/d_{yz}$ bands are reduced from a factor of 3--4 for FeSe ($T_c \approx 9$\,K) to 1.5--2 for FeS ($T_c \approx 4$\,K).

\begin{figure*}[htbp]
	\centering
\includegraphics[trim={0cm 0cm 0cm 0cm}, width=0.85\linewidth,clip=true]{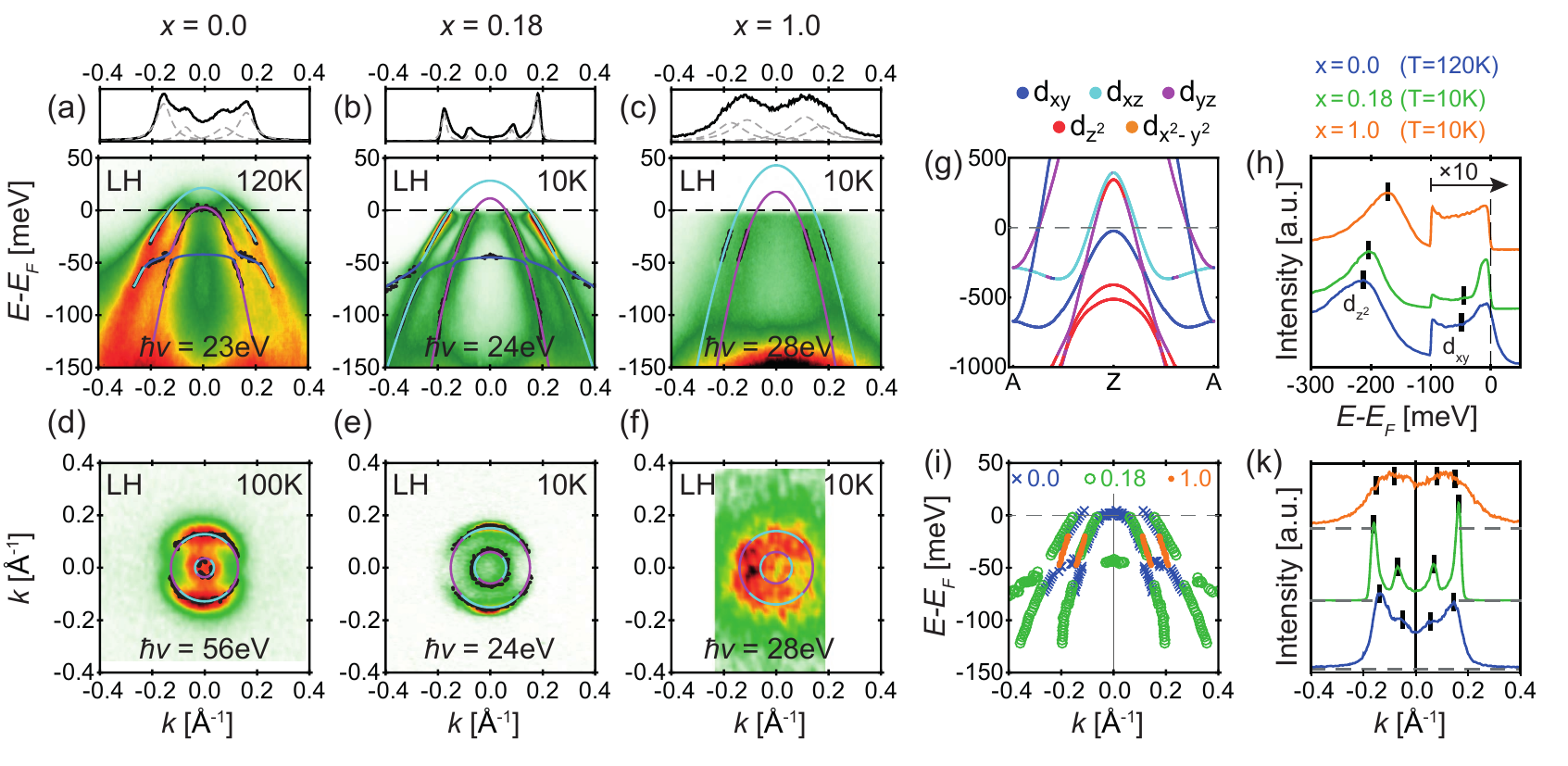}
	\caption{\textbf{The hole bands of tetragonal FeSe$_{1-x}$S$_x$ ($x=0, 0.18, 1$).}
 The ARPES spectra for  a) FeSe at 120~K, b) $x$=0.18 at 10~K and c) FeS at 10~K
 centered at high  symmetry $Z$-point.
 d)-f) The corresponding maps for the same compounds, as above.
 g) Band structure calculations for FeS using experimental parameters.
h) Energy distribution curves (EDCs) and k) momentum distribution curves (MDCs) for the three compounds. i) Extracted peak positions from fits to the MDCs from a)-c) for the hole bands  at the $Z$-point.}
	\label{fig1}
\end{figure*}

 {\it Experimental details}
 FeSe$_{1-x}$S$_x$ single crystals with $x$=0 and $x$=0.18  were grown by the KCl/AlCl$_3$ chemical vapour
transport method \cite{Chareev2013,Bohmer2016}.
FeS and other single crystals with $0.5 \leq x \leq 1$ were grown by the
hydrothermal method, using K$_{0.8}$Fe$_{1.6}$(Se$_{1-x}$S$_x$)$_2$ precursors \cite{Lai2015}.
ARPES measurements were performed at
the I05 beamline at the Diamond Light Source \cite{Hoesch2017},
%Single crystal samples were cleaved in-situ at a pressure lower than $2 \times 10^{-10}$~mbar
%at low temperatures.
 using horizontally and vertically linearly-polarised synchrotron light (LH and LV) between 20 and 120~eV,
%employing a Scienta R4000 hemispherical electron energy analyser
with $\approx $6 to 19 ~meV resolution.
Band structure calculations for FeS were performed with Wien2K using GGA, spin-orbit coupling and experimental lattice parameters,
($a$ = 3.6802(5)~\AA, $c$ = 5.0307(7)~\AA~ and $z_S$ = 0.2523 \cite{Lai2015}).

{\it Hole bands of tetragonal FeSe$_{1-x}$S$_x$.}
Fig.~\ref{fig1}(a)-(c) compares the hole band dispersions at the top of the Brillouin zone, centered at the $Z$ point, in the tetragonal phase of FeSe at $120\,$K ($T>T_s$) with those of $x$=0.18  and FeS at $10\,$ K. 
The photon energies corresponding to high-symmetry points along $k_z$ were established by analysis of the $d_{z^2}$ intensity well below $E_F$, shown in Fig.SM1 in the Supplemental Material (SM).
Despite the significant amount of sulfur substitution in $x$=0.18, the linewidths of the band dispersions in the ARPES spectra remain narrow due to the
high quality of these crystals, that also allows quantum oscillations to be observed \cite{Coldea2016}. This is in contrast to the much broader ARPES spectrum of FeS, shown in Fig.~\ref{fig1}(c), likely caused by the larger degree of disorder in crystals grown by the hydrothermal method with residual resistivity ratios varying between 5 to 17 (Fig.~\ref{fig3}(e)).

Band dispersions and the Fermi surface maps in Fig.~\ref{fig1} show that the high-$T$ band structure of FeSe and the low-$T$ band structure of $x$=0.18 are very similar, confirming the absence of the nematic state for $x$=0.18 at 10~K. The shape of the Fermi surface is isotropic in the $k_x-k_y$ plane for all three compositions (Fig.~\ref{fig1} (d)-(f)), in contrast to the elliptical Fermi surface found in the nematic phase of FeSe \cite{Watson2016} and for $x \leq 0.15$ \cite{Watson2015c}.
Two hole-like dispersions cross the Fermi level close to the $Z$ point, separated only by the spin-orbit coupling estimated  $\sim 20$\,meV in FeSe \cite{Watson2015a,Watson2017c}.

For a quantitative analysis of the band structure, band positions were extracted by performing simultaneous constrained Lorentzian fits to the momentum distribution curves (MDC) for different light polarizations at a fixed energy, shown in Fig.~\ref{fig1}(i) and (k). For FeS, best fits were obtained also using two hole-like bands at the Fermi level, as expected from the band structure calculations (Fig.~\ref{fig1}(g)), even though the two bands are harder to separate (see also Fig.SM3 in SM). We find a measurable increase of the $k_F$ values and the Fermi surface areas with increasing S substitution (Fig.~\ref{fig1}(i)), in agreement with the trends found in quantum oscillations up to $x\approx 0.19$ \cite{Coldea2016}.

One unusual feature of the electronic structure of FeSe is the existence of a small 3D hole pocket centered only around the $Z$ point at 100~K (Fig.~\ref{fig1}(a) and (d)). This innermost hole band is pushed below the Fermi level at low temperatures, due to the combined effects of orbital order and spin-orbit coupling \cite{Watson2015c,Fernandes2014b}. As orbital ordering is reduced with S substitution, this small 3D pocket reappears at $Z$ for $x \sim 0.11$ at low temperatures \cite{Watson2015c}, consistent with our observations for $x = 0.18$ (Fig.~\ref{fig1}(b) and (e) and Fig.SM3). However, in  FeS, due to the significant increase in  bandwidths, we find that this pocket has become two-dimensional, as evidenced by two bands crossing the Fermi level both at the $\Gamma$ and $Z$ point (Fig.SM3).

\begin{figure*}[htbp]
	\centering
\includegraphics[trim={0cm 0cm 0cm 0cm}, width=0.85\linewidth,clip=true]{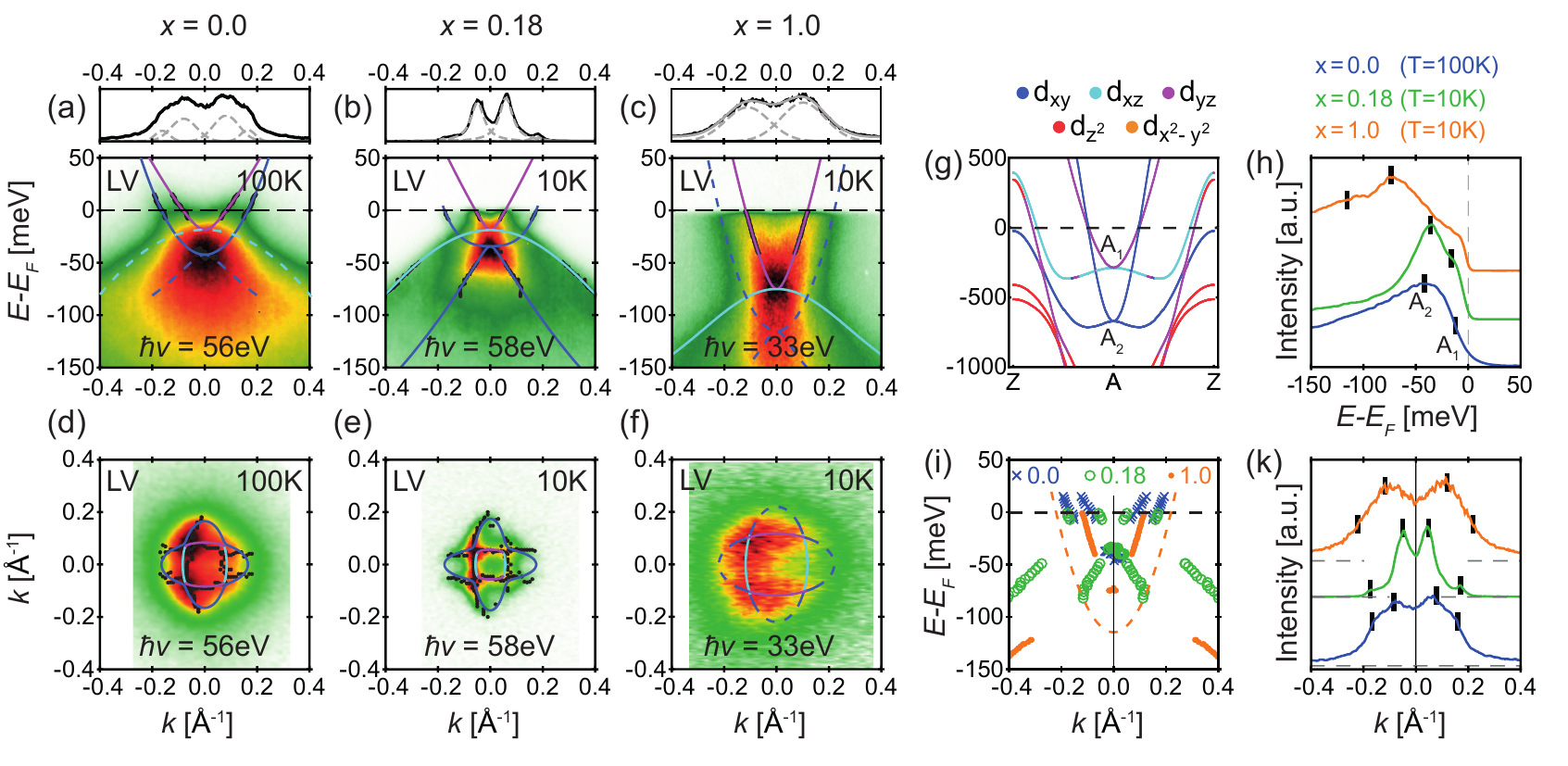}
	\caption{\textbf{The electron bands of tetragonal FeSe$_{1-x}$S$_x$ ($x=0, 0.18, 1$).}  a-c) ARPES intensity plots of the band structure through the $A$ point
and d)-f) the corresponding maps for the three compounds, as in Fig.~\ref{fig1}.
 g) Band structure calculations for FeS using experimental parameters.
h) EDCs and k) MDCs for the three compounds. i) Extracted peak positions from fits to the MDCs from a)-c).}
	\label{fig2}
\end{figure*}

Next, we compare the change in the electronic correlations as a function of S substitution. The strongest renormalizations are expected for bands with $d_{xy}$ character, but in ARPES, they are notoriously difficult to observe due to matrix element effects. However, their dispersions can be revealed due to band hybridization caused by the spin-orbit coupling effects \cite{Watson2015a,Watson2016,Fanfarillo2016,Fedorov2016}. This allows us to identify the $d_{xy}$ hole band in FeSe and FeSe$_{0.82}$S$_{0.18}$, and we find it significantly pushed below the Fermi level ($ \sim 50$~meV), in contrast to band structure calculations where it crosses the Fermi level (see Fig.~SM2). In FeS, the $d_{xy}$ band is not resolved due to disorder effects, as found in other iron-based superconductors \cite{Ye2014}, rather than due to the increase in correlations when it can become incoherent, as observed in FeSe$_{x}$Te$_{1-x}$  \cite{ZKLiu2015}.

In FeSe, the $d_{xy}$ band renormalization is rather large (a factor 7-9), in contrast to the $d_{xz}$/$d_{yz}$ band renormalization (a factor 3-4) \cite{Watson2015a,Maletz2014} and we find that they do not change significantly when comparing to $x$=0.18, shown in Fig.~\ref{fig1}(i).  However, for FeS we extract a significantly reduced band renormalization of 1.7(1) for the $d_{xz}$/$d_{yz}$ bands, reflecting moderate electronic correlations for FeS with a low $T_c \sim 4$~K. An additional band present in FeS is the $d_{z^2}$ band, which lies closer to the Fermi level ($\approx 150\,$meV) as compared with DFT (350~meV), also suggestive of finite correlations effects in FeS
(renormalized by a factor $\approx 2$ from the $k_z$
dependence in Fig.SM1 in SM).
Furthermore, the Fermi velocities $v_F$ extracted from the band dispersion slopes (Fig.~\ref{fig1}(i) and Fig.~\ref{fig3}(d)) significantly increase  from FeSe towards FeS,  whereas the quasiparticle effective masses, $m^\star$,  of the outer hole-like bands decrease from 3-4~$m_e$ for $x=0.18$ to 1-2~$m_e$ for FeS. These findings agree
with the reduction of the effective masses detected in  quantum oscillations studies in FeSe$_{1-x}$S$_x$ (outside the nematic phase) \cite{Coldea2016} and in FeS \cite{TerashimaFeS}.

{\it Electron bands of tetragonal FeSe$_{1-x}$S$_x$.}
Fig.~\ref{fig2}(a)-(c) compares the evolution of the band structure at the A point in the tetragonal phase of FeSe at 100~K, and of $x$=0.18 and FeS at 10~K. As for the hole-like bands at the Z point, the ARPES spectra of FeSe and FeSe$_{0.82}$S$_{0.18}$ are very similar, confirming that for $x \sim 0.18$, the Fermi surface deformation observed in the nematic state of FeSe is completely suppressed \cite{Watson2015a,Watson2015c}. The spectra of all samples display two electron-like bands crossing the Fermi level, but they are much harder to separate for FeS (Fig.SM3). For FeS we use a single band fit to the MDCs in Fig.\ref{fig2}(k), whereas the outer electron band size with $d_{xy}$ character 
is affected by matrix elements and disorder effects.
 Fermi surface maps in Fig.~\ref{fig2}(d-f)  display a four-fold symmetric shape, with small differences between the inner electron-like Fermi surface pocket between $x$=0.18 and FeSe at 100\,K, whereas a significant expansion is detected for FeS (Fig.~\ref{fig2}(f)).

At the corners of the tetragonal Brillouin zone, there are two degenerate states, $A_1$ and $A_2$ (Fig.\ref{fig2}(g)),
which are the bottom of the inner and outer electron bands and are not split by the spin-orbit interaction \cite{Fernandes2014b}. The increased separation between these states upon cooling through the nematic transition has caused a significant debate about the origin of the nematic phase \cite{Fernandes2014b,Watson2016,Watson2017c,Evtushinsky2016_arxiv}.
Here we find the bottom of the inner electron band is $\approx 19$\,meV below the Fermi level  for $x$=0 and $x$=0.18, with slight variation for the outer electron band ($ \approx 42$\,meV for $x$=0 and $ \approx 34$\,meV for $x$=0.18). Notably in FeS, these two degenerate states are significantly lower in energy compared with the other two compositions ($\approx 70 $\,meV and $\approx 120 $\,meV, respectively), a direct consequences of the increased bandwidths
(identified from the energy distribution curves (EDC) shown in Fig.~\ref{fig2}(h)).
 This extended bandwidth, in conjunction with the equally significant increase of the Fermi velocity (Fig.~\ref{fig2}(i) and Fig.~\ref{fig3}(d)) and a decrease of the quasiparticle effective masses, highlight the significant reduction in the electronic correlations in FeS.

 \begin{figure*}[htbp]
	\centering
	\includegraphics[trim={0cm 0.1cm 0.1cm 0.1cm}, width=0.80\linewidth,clip=true]{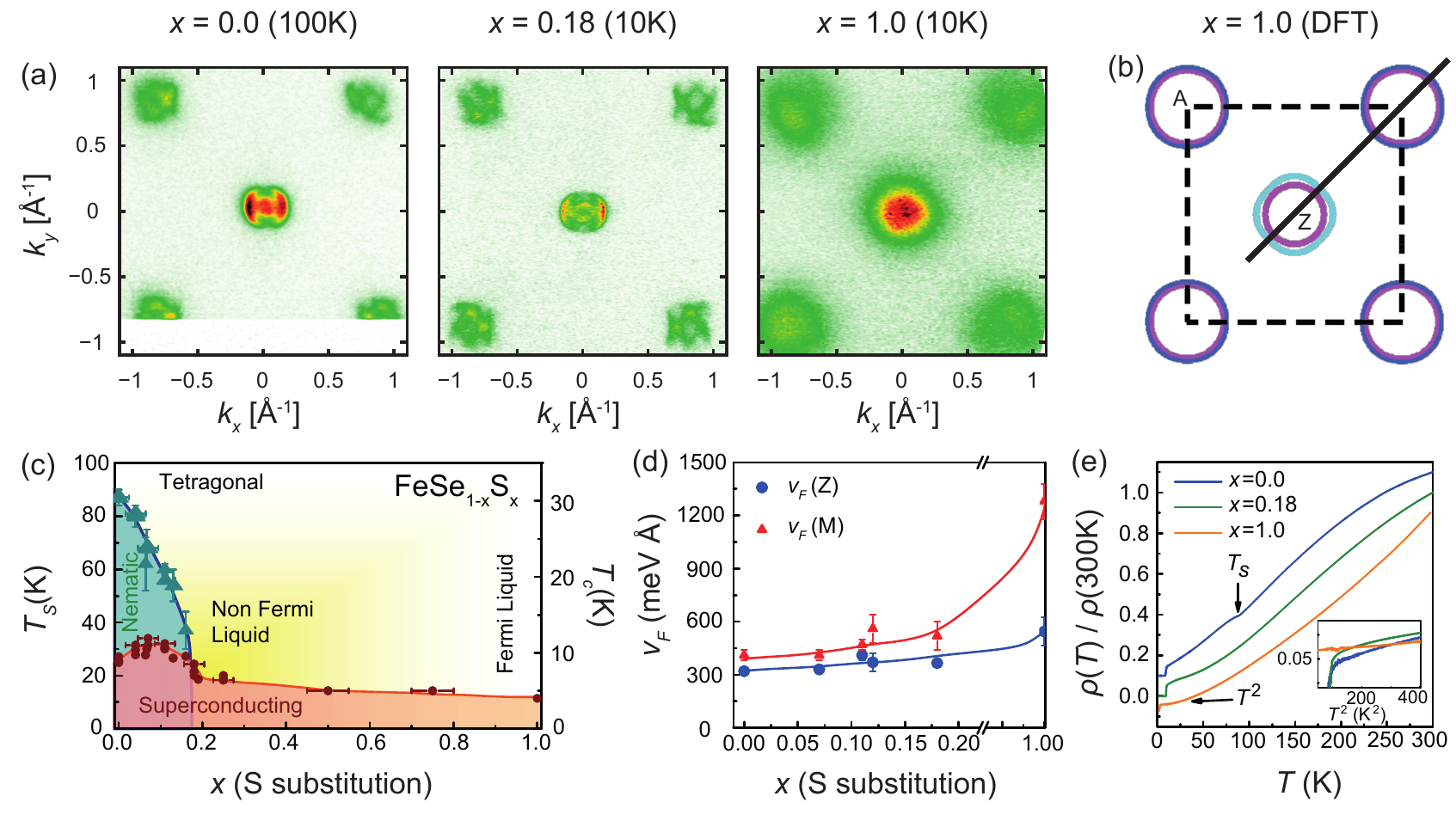}
	\caption{\textbf{Phase diagram of FeSe$_{1-x}$S$_x$  and the suppression
			of electronic correlations by S substitution.} a) The ARPES map of the
		high-symmetry cut through top of the Brillouin zone  
		 for the tetragonal phase of FeSe$_{1-x}$S$_x$ ($x$=0, 0.18, 1) at 56-69~eV, together with a calculated slice for FeS in b). The solid line indicates the cuts used during ARPES experiments.
		c) Proposed phase diagram of FeSe$_{1-x}$S$_x$, including different transport and thermodynamic measurements from Refs.\cite{Coldea2016,Mizuguchi2009,Abdel-Hafiez2015}.  d) The evolution of the Fermi velocities as a function of chemical pressure.
		e) The temperature dependence of resistivity of FeSe$_{1-x}$S$_x$ ($x$=0, 0.18, 1) showing strong deviations
		of resistivity from $T^2$ Fermi liquid behaviour for low sulfur substitution. The data are renormalized to the room temperature
		values and shifted for clarity.}
	\label{fig3}
\end{figure*}

{\it The phase diagram of FeSe$_{1-x}$S$_x$}
together with the evolution of the Fermi surface in the tetragonal phase from FeSe to FeS is shown in
Fig.\ref{fig3}(c) and Fig.\ref{fig3}(a), respectively.
While the size of the quasi-two dimensional Fermi surface increases with chemical pressure, the most important change is the increase in Fermi velocities
(and bandwidths) (Fig.\ref{fig3}d), which reflects the reduction of the electronic correlations. These findings
agree with the reduction of the effective masses determined from quantum oscillations in FeSe$_{1-x}$Se$_x$ outside the nematic phase \cite{Coldea2016,Reiss2017pressure} and FeS \cite{TerashimaFeS,Man2016}. Furthermore, the low  temperature resistivity shows a $T^{2}$ Fermi-liquid-like behavior for FeS, in contrast to the other compositions closer to the nematic phase, as shown in Fig.\ref{fig3}(e) and also reported in Ref.\cite{Urata2017}.
The low-$T_c$ superconductivity in FeSe$_{1-x}$S$_x$ has a small dome inside the nematic region, being gently suppressed towards FeS (Fig.\ref{fig3}(c)).
This behavior is in contrast to FeSe under applied pressure \cite{Sun2016pressure} or {\it in-situ} K dosing \cite{Wen2016}, where superconductivity is enhanced once the nematic phase is suppressed, with an additional magnetic phase being stabilized under pressure \cite{Sun2016pressure,Terashima2015,Medvedev2009}.

Our results on the electronic structure
of FeS are in good agreement with a recent ARPES study \cite{Miao2017}.
Quantum oscillations in FeS
reported only small frequencies below 200~T
\cite{TerashimaFeS}, a factor 2.5 smaller than the
smallest area of the inner hole band predicted by
band structure calculations \cite{Coldea2016}.
Our ARPES data do not reveal the presence of such a small band 
(with a $k_F \sim 0.0780 $\AA$^{-1}$), but the $d_{xy}$
is not visible due to the matrix effects or impurity line broadening.
This $d_{xy}$ band is predicted to lie very close
to the Fermi level in FeS (Fig.~\ref{fig1}f).
Furthermore, due to the complex
de-intercalation procedure to prepare FeS, other
byproducts could form \cite{Kirschner2016}.
Recently, a quantum oscillations study
suggested that FeS has a 3D Fermi surface \cite{Man2016},
not supported by the current ARPES studies.

As FeS remains in the tetragonal phase
and the electronic correlations are reduced,
one would expect a better agreement between the
experimental and calculated Fermi surface of FeS (see Fig.\ref{fig3}b).
However, we find that the Fermi surface areas
and the quasiparticle masses of FeS are still a factor $\sim 2$ smaller than
predicted by DFT calculations (Fig.\ref{fig3}b and Fig.SM2).
This band shrinking thus also manifests in FeS, but 
is weaker than in FeSe \cite{Watson2015a}.
Furthermore, FeS is reminiscent of 
other iron-based superconductors with a low $T_c$, LaFePO and LiFeP,
where the renormalization effects extracted from quantum oscillations were rather moderate ($\approx 2$)
\cite{Coldea2008,Putzke2012}. Interestingly,
all these end member compounds,
LaFePO and LiFeP and FeS, display nodal superconductivity \cite{Fletcher2009,Hashimoto2012,Xing2016,Yang2016}
and the pnictogen and chalcogen position is closer to the iron planes
compared to their isoelectronic sister-compounds.
These trends have been
captured theoretically by Kuroki {\it et al.} \cite{Kuroki2009},
where the height of the pictogen
acts as a switch between high-$T_c$ nodeless and low-$T_c$ nodal pairings
and that superconductivity is suppressed
once the lattice constants are reduced, as in the case
of FeS.
Substituting smaller S ions onto the Se site shrinks
the unit cell \cite{Lai2015,Mizuguchi2009}, decreases the Fe chalcogen bond lengths
and brings the chalcogen closer to the iron planes.
This would result in a greater orbital overlap causing an increase in the bandwidth and the degree
of electronic correlations will reduced significantly, like in FeS.

{\it Summary}. Our high-resolution ARPES study on FeSe$_{1-x}$S$_x$ single crystals
reveal the suppression of the electronic correlations, 
demonstrated by the  increase in Fermi velocities and bandwidth,
while the superconductivity is weakened away from the nematic phase.
The chemical pressure effects in FeSe$_{1-x}$S$_x$ lead to the increase in the size of the quasi-two dimensional
Fermi surface, however,  
its size still remains smaller than predicted from first principle band structure calculations.
Our results suggest that electronic correlations may be important
for enhancing superconductivity in iron-based superconductors and chemical pressure
offers an ideal tuning parameter to  control them.

{\it Acknowledgements}
We thank Moritz Hoesch for technical support.
This work was mainly supported by  EPSRC  (EP/L001772/1,  EP/I004475/1,  EP/I017836/1).
A.A.H. acknowledges the financial support of the Oxford Quantum
Materials Platform Grant (EP/M020517/1).
We  thank  Diamond  Light  Source  for  access  to  Beamline
I05 (proposal number SI15471) that contributed to the results
presented  here.  The authors would like to acknowledge the use of the
University of Oxford Advanced Research Computing (ARC)
facility in carrying out part of this work.
A.I.C. acknowledges an EPSRC Career Acceleration Fellowship (EP/I004475/1).

\bibliography{FeSeS_bib_may17}

\end{document}